\documentclass{egpubl}
\usepackage{egsr2026}
\usepackage{colortbl}
\usepackage{soul}
\usepackage{wrapfig}
\usepackage{amsmath}
\usepackage{amsfonts}

\definecolor{tabfirst}{rgb}{1, 0.7, 0.7}   
\definecolor{tabsecond}{rgb}{1, 0.85, 0.7} 
\definecolor{tabthird}{rgb}{1, 1, 0.7}     
 
\SpecialIssuePaper         

\CGFccby

\usepackage[T1]{fontenc}
\usepackage{dfadobe}  

\biberVersion
\BibtexOrBiblatex
\usepackage[backend=biber,bibstyle=EG,citestyle=alphabetic,backref=true]{biblatex} 
\electronicVersion
\PrintedOrElectronic

\ifpdf \usepackage[pdftex]{graphicx} \pdfcompresslevel=9
\else \usepackage[dvips]{graphicx} \fi

\usepackage{egweblnk} 

\title{Lighting-Consistent Object Transfer Across Radiance Fields}

\author[N. Violante et al.]
{\parbox{\textwidth}{\centering N.\,Violante$^{1}$\orcid{0009-0000-3169-8075},
        G.\,Kopanas$^{2}$\orcid{0009-0002-5829-2192},
        L.\,Franke$^{1}$\orcid{0000-0001-8180-0963},
        J.\,Philip$^{3}$\orcid{0000-0003-3125-1614}
        and G.\,Drettakis$^{1}$\orcid{0000-0002-9254-4819}
        }
        \\
{\parbox{\textwidth}{\centering $^1$Inria, Université Côte d'Azur, France
         $^2$Google DeepMind, United States
         $^3$Eyeline Labs, United Kingdom
       }
       }}

\begin{document}
\teaser{
 \includegraphics[width=\linewidth]{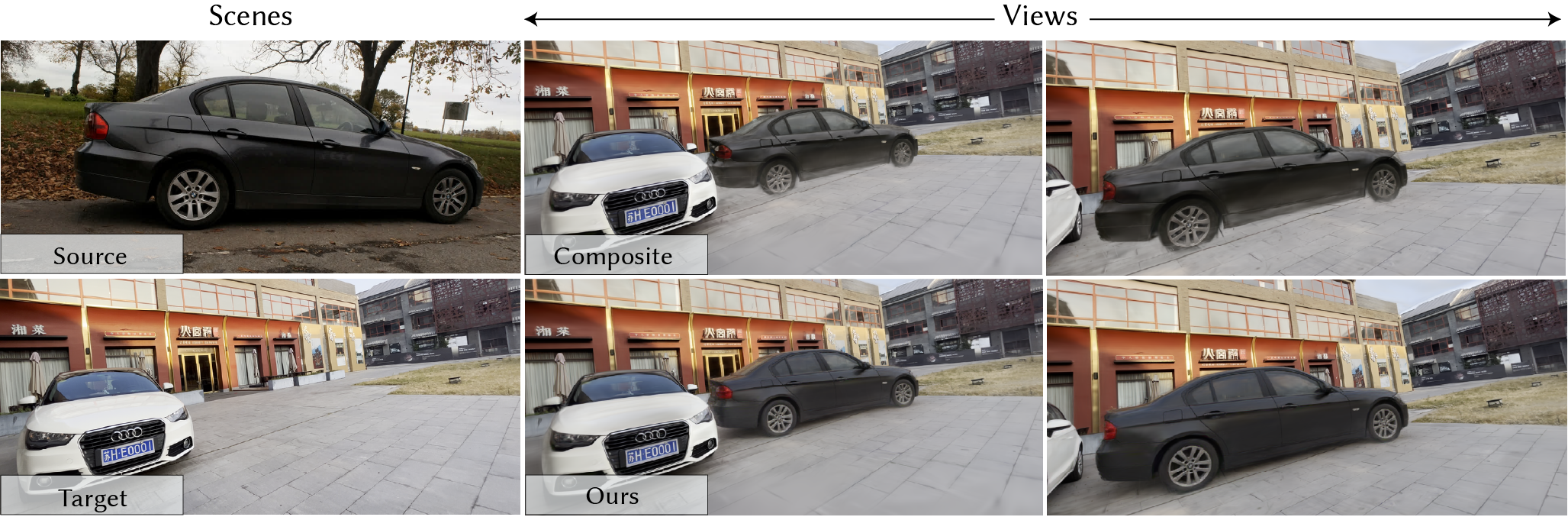}
 \centering
  \caption{
  We transfer an object from one captured 3DGS scene into another and harmonize its lighting so that shadows and reflections match the new environment. Naively compositing the source object (the car, from Ref-NeRF~\cite{verbin2021ref-nerf}) into the target scene (from DL3DV~\cite{ling2024dl3dv}) leaves it inconsistently lit, with missing shadows and wrong reflection highlights (upper row). Our method harmonizes the object and produces a multi-view consistent result (lower row).
  }
\label{fig:teaser}
}

\maketitle
\begin{abstract}
3D Gaussian Splatting (3DGS) is widely used to capture and render real scenes. Compositing objects from one capture into another has applications in many domains, such as VFX, architecture and interior design, or marketing. However, extracting an object from a source scene and naively pasting it into a target scene will fail to produce realistic results due to the different lighting conditions between the two scenes. 
To address this problem, we introduce a diffusion model that harmonizes naively composited images with inconsistent lighting. The model is trained with a heterogeneous dataset of image pairs (inconsistent composite input, consistent output), combining synthetic, generated, and real data. Our complete 3D solution allows a user to extract an object from the source scene and composite it into the target scene. From this, the (inconsistent) views of the target scene with the composite object are rendered. Our diffusion model harmonizes each one of these views, which are finally consolidated in a 3DGS representation with a post-optimization step. 
Our method provides visually compelling results, making object transfer between 3DGS easy to use and significantly improving quality compared to previous methods.
\\

\printccsdesc   
\end{abstract}  

\newpage

\section{Introduction}

3D Gaussian Splatting (3DGS)~\cite{kerbl_3Dgaussians_2023} allows easy capture and real-time free-viewpoint navigation for real scenes. Unfortunately, such representations are hard to edit, since they entangle materials and lighting as they appear at capture time. Compositing scenes and objects captured independently has many real-world applications ranging from VFX and architectural visualization to interior design and marketing. But naively compositing scenes by copying Gaussian primitives from a source scene to a target scene leads to unrealistic results because of the lighting inconsistencies between the two scenes (Fig.~\ref{fig:teaser}, top row). We introduce a method that allows a user to extract an object from one captured 3DGS scene, interactively place it into another scene in 3D, and produce a composited scene with consistent lighting.

Traditional solutions to the composition problem involve modeling the geometry and materials of the objects, and rendering using estimated light from the captured target scene; this is technically demanding and costly, requiring both expert 3D modeling and reliable lighting estimation. Previous automated methods have tried to harmonize the illumination of an inserted object for the 2D image case~\cite{niu2025makingimagesrealagain}. Recently, diffusion models have been used as priors to allow relighting of images, producing consistent highlights and shadows~\cite{jin2024Neural,Choi:2025:SSI,magar2025LightLab}, and also to perform intrinsic image decomposition and editing~\cite{zeng2024RGB-X,kocsis2023intrinsic, careaga_intrinsic_2023, careaga_physically_2025}. Lifting such 2D image methods to 3D requires making the result consistent over all input views used for capture. This is a hard task due to the inverse and ambiguous nature of the problem combined with the lack of hyper-realistic 3D datasets. Diffusion models can also relight captured 3D objects in the simplified case of distant illumination~\cite{jiangGaussianShader3DGaussian2024,liangGSIR3DGaussian2024}, but ignore cast shadows and foreground-background interactions. These models have been used to relight more general scenes~\cite{poirier_GSrelighting_2024}, but composition and harmonization across captures are not feasible with such methods. 

We thus present DOT3D, \emph{Diffusion Object Transfer in 3D,} 
a method to transfer objects from a source scene into a target scene.
DOT3D allows the user to interactively extract an object from a source 3DGS scene, place it at the desired location in a target 3DGS environment, and harmonize the object to match its new environment. To harmonize the object in the target scene, we first render each input view of the naively composited scene and use a fine-tuned diffusion model to harmonize the \emph{inconsistent lighting} in these renders. We then consolidate the independently harmonized views in 3D so they are multi-view consistent, producing a fully harmonized 3DGS scene.

In practice, this process faces two challenges: 1) fine-tuning a diffusion model to harmonize images requires a high-quality dataset, and 2)
ensuring multi-view consistency of the individually harmonized images.
To address the first challenge, we create a heterogeneous dataset that combines three sources: synthetic data with ground-truth values rendered using global illumination, paired multi-illumination data generated by a high-quality diffusion model, and paired real data showing scenes with and without an object, augmented with a relighting network.
For the second challenge, we propose a 3DGS post-optimization to consolidate the individually harmonized views into a consistent 3DGS representation.

In summary, our main contributions are: 
\begin{itemize}
  \item A diffusion model specialized for 2D lighting harmonization of image object insertion, trained with a heterogeneous dataset that includes multiple sources of synthetic, generated, and real data.
  \item A 3DGS post-optimization for multi-view consolidation of independently harmonized views to create a consistent scene.
 \item An interactive pipeline allowing users to extract an object from a source scene, and insert it into a target capture with consistent lighting.
\end{itemize}

Our results on a variety of commonly used datasets demonstrate that we achieve plausible and consistent object transfer between captured real scenes. Our code and data are available at \url{https://repo-sam.inria.fr/nerphys/dot3d}

\section{Related Work}

\textbf{Novel View Synthesis with Radiance Fields.}
Neural Radiance Fields (NeRF)~\cite{mildenhall2020nerf, barron2021mip-nerf, barron2022mip-nerf360} represent a scene with a multi-layer perceptron (MLP) that maps position and viewing direction into volumetric density and view-dependent color. The colors along pixel rays are integrated to produce the rendered image. Despite using acceleration structures~\cite{muller2022instant, fridovich-keilPlenoxelsRadianceFields2022}, rendering each pixel involves marching along a ray and querying the MLP hundreds of times, making rendering slow. To overcome this limitation, 3D Gaussian Splatting (3DGS)~\cite{kerbl_3Dgaussians_2023} represents a scene with a set of Gaussian primitives that can be efficiently rasterized on the GPU, allowing real-time rendering at over 100 FPS. Since its introduction, 3DGS has been extended to address some of its limitations, notably high-frequency reflections~\cite{ye3DGaussianSplatting2024,kourosRGSDRReflectiveGaussian2025}, and has shown promising development for relighting and inverse rendering in the simplified object-centric case~\cite{jiangGaussianShader3DGaussian2024,liangGSIR3DGaussian2024}.

\textbf{Radiance Field Segmentation.} To extract an object from one scene and insert it into another, we must identify which Gaussians belong to the object. Most methods extend the Gaussian primitives with additional features, and then optimize these features to match 2D masks of the object across different views~\cite{siddiqui_panoptic_2023, qin_langsplat_2023, qiu_featuresplatting_2024,Zhou_Feature3DGS_2024, lee_rethinking_2024, ye_gaussian_2023, Zhai_2025_PanoGS}. During optimization, contrastive methods check whether two pixels belong to the same mask. This avoids tracking mask correspondences across views, but makes optimization slow~\cite{gu_egolifter_2024, kim_garfield_2024, choi_clickGaussian_2024, cen_segment_2024, ying_omniseg3d_2024, SplatAndReplace2025Violante}. Since we extract a single object, we optimize the features of the primitives to match a binary mask using a simple binary classification loss, making optimization faster.

\textbf{Light Control with Generative Models.} To insert an object into a new environment, we must relight the object and cast proper shadows and reflections to match the new environment. Early approaches based on GANs~\cite{karras2019style, karras2020stylegan2} edit the lighting of generated scenes via latent space manipulation~\cite{harkonen2020ganspace, bhattad2023stylegan_knows, wang2022stylelight}. However, training GANs is prone to instabilities and mode collapse~\cite{bau2019seeing}. Diffusion models~\cite{sohl2015deep, hoDenoisingDiffusionProbabilistic2020, dhariwalDiffusionModelsBeat2021, song2019generative, song2020score} are more stable to train and offer unprecedented high-quality text-to-image synthesis at large scale~\cite{rombachHighResolutionImageSynthesis2022, ramesh2022hierarchical}. These models require costly training, which has motivated fine-tuning and control methods that adapt pre-trained models to novel tasks~\cite{huLoRALowRankAdaptation2021, yeIPAdapterTextCompatible2023, zhang_controlnet_2023}.

Building on this progress, diffusion models have been widely adopted by the graphics and vision communities, and in particular, they have been extended for lighting manipulation tasks on images. Recent work has enabled direct control over light sources~\cite{poirier_GSrelighting_2024, magar2025LightLab}. Another line of work, inspired by 3D modeling workflows, employs intrinsic decomposition methods~\cite{zeng2024RGB-X, kocsis2023intrinsic, luo2024IntrinsicDiffusion} to extract channels such as albedo, roughness, normals, and shading, which in turn enable editing through direct manipulation of these channels. Applications include object insertion, removal, and relighting~\cite{lyu2025IntrinsicEdit, zhang2025ZeroComp, careaga_intrinsic_2023}, but these approaches require precise edits on multiple channels, and are restricted to the 2D case. 

Careaga and Aksoy~\cite{careaga_physically_2025} propose an intrinsic-based approach for 2D relighting with an intermediate mesh estimation to control lighting. Recently, initial efforts have been made to consolidate 2D intrinsic decomposition into 3D~\cite{kocsis2025intrinsic,langsteiner2025matspray}, but with restrictive assumptions, e.g., complete mesh available, or isolated objects only lit by environment maps.

\textbf{Composition and Inpainting.}
Composition methods aim to integrate objects given a coarse placement~\cite{NeRFshop23} but often struggle to preserve object identity and pose~\cite{song2022ObjectStitcha, yang2022Paint, chen2024AnyDoor}.
Similar in spirit to our method, Nicolet et al.~\cite{NPD20a} use a relighting network to harmonize lighting on images and rely on Unstructured Lumigraph~\cite{2001lumigraph} with meshes for novel view synthesis, thus restricting quality. ObjectDrop~\cite{winter2024ObjectDrop} enables precise placement and adds shadows and reflections to an inserted object. But unlike our work, it is restricted to 2D and does not account for the lighting of the inserted object, which generates inconsistent illumination when the object is taken from a scene with noticeable differences in lighting conditions.

Diffusion-based inpainting methods enable the generation of content within a user-specified region of an image~\cite{lugmayr2022repaint}. To build a dataset with different lighting conditions, ControlCom~\cite{zhang2023controlcom} applies only image-space operations (jitter, saturation, brightness) and is restricted to 2D. SpotLight’s~\cite{fortier2024spotlight} data pipeline lacks generated data and real data, and the method needs guidance from shadow images.

In the 3D context, recent inpainting methods for 3DGS often focus on object removal~\cite{huang20253D, liu2024InFusion}. For object insertion, D3DR~\cite{skorokhodov2025D3DR} personalizes a diffusion model on a few images of the object~\cite{ruiz2023dreambooth}, but struggles to obtain a realistic integration of the object with its surroundings. MVInpainter~\cite{cao2024MVInpainter} uses a multi-view network to inpaint multiple views.
MV-CoLight~\cite{renMVCoLightEfficientObject2025a} proposes two feed-forward transformers for harmonization: one for individual images, and another for Gaussian primitives. The method focuses on low-resolution images ($256\times256$) and scenes with few images, typically between 6 and 16, covering limited points of view. In contrast, our method handles full scenes with hundreds of high-resolution images.

\textbf{Diffusion for Novel-view and 3D Reconstruction.} Both image and video diffusion models demonstrate outstanding generation capabilities. They are trained on massive datasets that have no equivalent in 3D, where data is scarce. Thus several works have explored the use of these pre-trained models to generate 3D content. CAT3D \cite{gaoCAT3DCreateAnything2024} leverages multi-view diffusion to generate a scene that is then baked into a 3DGS representation, while other approaches \cite{renGEN3C3DInformedWorldConsistent2025, yuTrajectoryCrafterRedirectingCamera2025} use camera-conditioned video models. Complex high-level editing and generation can also be performed using the priors from diffusion models \cite{haque2023instruct}. While designed for generative tasks, the natural image and video priors of these models have also been used to improve 3D reconstruction and novel view synthesis. The priors can help reconstruction in underconstrained regions with Gaussian artifacts~\cite{liu3DGSEnhancerEnhancingUnbounded2024,wuDifix3DImproving3D2025}, in sparse view settings \cite{wuReconFusion3DReconstruction2024}, and when content is missing \cite{wuGenFusionClosingLoop2025}. Such models can also be fine-tuned to remove Gaussian-like artifacts as a post process, allowing very high resolution in constrained cases~\cite{DEGS}.
\begin{figure}
    \centering
    \includegraphics[width=\linewidth]{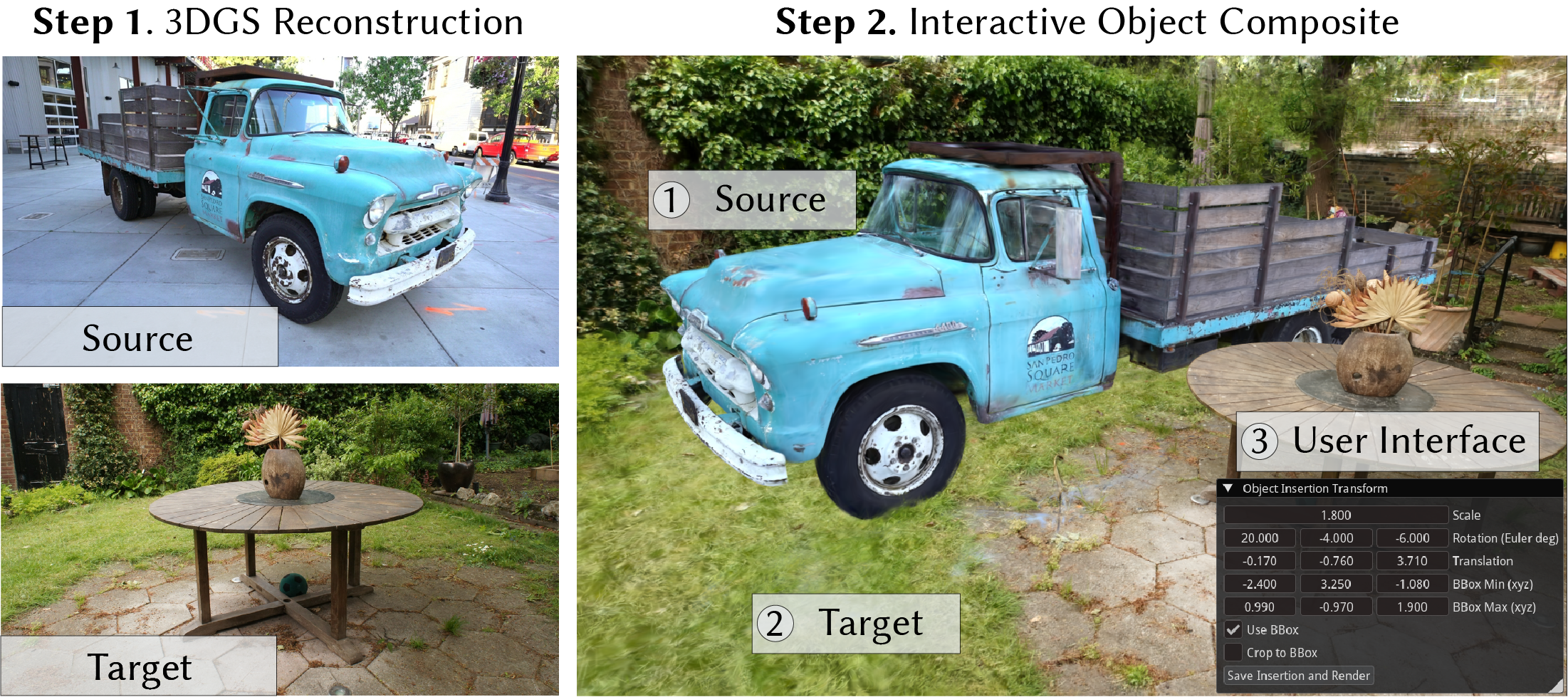}
    \caption{\textbf{Object Composition Interface.} We propose a user interface for object composition across radiance fields. The user loads the extracted object from the source scene, inserts it into the target scene, and can control its final position, orientation, and scale within the target scene. Composite images are rendered using the camera viewpoints from the target scene.}
    \label{fig:interface}
\end{figure}
\begin{figure}
    \centering
    \includegraphics[width=\linewidth]{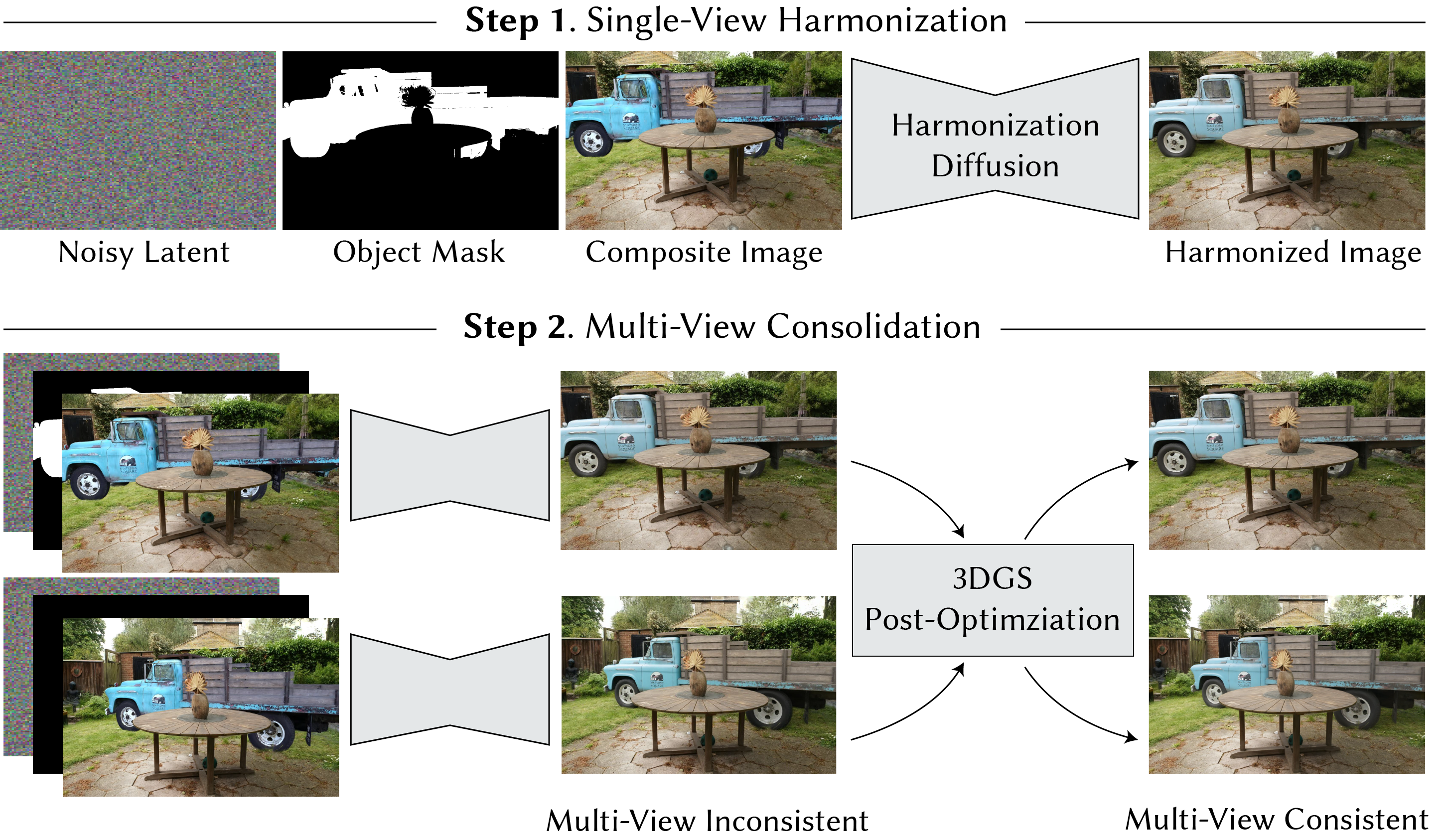}
    \caption{\textbf{Method Overview}. \textit{Step 1: Single-View Harmonization:} Conditioned on the composite image and the object mask, our harmonization model transforms a noisy latent into a harmonized image with plausible lighting, shadows, and reflections. \textit{Step 2: Multi-View Consolidation:} Our 3DGS post-optimization consolidates the independent harmonized single-views into a multi-view consistent representation.}
    \label{fig:overview}
\end{figure}

\section{Overview}
Our method transfers an object from a source scene into a target scene, harmonizing it to match its new environment. An overview of our method is presented in Fig.~\ref{fig:interface} and Fig.~\ref{fig:overview}.

We first reconstruct a 3DGS representation of the source scene and extract an object by lifting 2D masks to the Gaussian primitives, enabling 3D segmentation. We then use an interactive interface to insert the object in a target 3DGS scene, with full control of the object's position, orientation, and scale (Fig.~\ref{fig:interface}). 

To harmonize each view of the target scene with the composited object, our diffusion model takes as input a render of the naively composited scene together with a binary mask of the inserted object, and produces a harmonized output image. In this output image, the object matches the target scene's lighting and is visually integrated with its surroundings, including secondary effects such as shadows and reflections. To train our diffusion model, we construct paired examples where the input shows the object under inconsistent lighting while the target output depicts the same object consistently inserted into the scene. This enables the model to learn how to transform lighting-inconsistent composites into harmonized images.

While this process yields harmonized images for individual views, the results are independent and thus not multi-view consistent.
To address this, we propose a post-optimization of the composited 3DGS representation. Starting from this composited scene, we optimize the colors of the Gaussian primitives with a perceptual loss, while keeping all other Gaussian attributes (geometry, opacity) constant.
This procedure enforces consistency while preserving fidelity, ultimately producing a consistent insertion of the object within the target scene.

\section{Interactive Object Composition}\label{sec:object_composition}
Our approach begins by reconstructing the source scene and extracting the object of interest from it. To this end, we obtain 2D binary masks of the foreground object using a pre-trained BiRefNet~\cite{zheng2024birefnet}, although other options are also well-suited for this task (e.g., SAM models~\cite{kirillov2023segany, ke_samhq_2023, ravi_sam2_2024}). We extend 3DGS for 3D segmentation by assigning to each Gaussian primitive an extra feature, alongside the color attributes, of dimension $d=1$, which can be efficiently rasterized on the GPU. To train these features, we rasterize them into per-pixel features and optimize a binary classification cross-entropy objective using the binary masks as targets.

We then segment the object in 3D from this source scene using a similarity threshold of $0.75$ on the binary features (after sigmoid activation), and transfer it into a target scene using a 3D interface that we designed for this purpose (Fig.~\ref{fig:interface}).
This interface enables users to adjust the object's position and orientation within the target scene, and also provides tools to refine the segmentation through bounding box selection to remove unwanted artifacts. 
This produces an \emph{inconsistent} composite of the object from the source scene into the target scene, which we later harmonize.

\section{Single-View Harmonization}\label{sec:relighting_method}

\newcommand{\Znoisy}{\mathbf{z}_t}
\newcommand{\Itgt}{\mathbf{x}^{\text{out}}}
\newcommand{\Isrc}{\mathbf{x}^{\text{comp}}}
\newcommand{\Imask}{\mathbf{M}}
\newcommand{\Ibg}{\mathbf{x}^{\text{bg}}}
\newcommand{\Ifg}{\mathbf{I}^{\text{incon}}}
\newcommand{\Zcomp}{\mathbf{z}_{\text{comp}}}
\newcommand{\Zmask}{\mathbf{z}_{\text{mask}}}

After compositing, the object is not yet properly integrated into the target scene due to the lack of consistent lighting. To address this, we propose a diffusion-based harmonization model to adapt the object’s illumination and its surroundings to fit in the target scene (Fig.~\ref{fig:overview}).
Our harmonization model takes as input an image of the composite object under mismatched illumination, 
and outputs a harmonized image where the object's lighting matches that of the target scene and is integrated with shadows and reflections, yielding a result that appears naturally captured rather than naively composited. Fine-tuned diffusion models, which exploit the rich data priors of the base model, are a powerful way to handle problems such as ours, but a critical component for success is the curation of a dataset for the specific task. 

To build our model, we follow DEGS~\cite{DEGS} and fine-tune
FLUX.1-schnell~\cite{labs2025flux1kontextflowmatching}, a state-of-the-art rectified-flow model~\cite{lipman2022flow} for text-to-image generation. FLUX.1-schnell produces high-quality results in only 4 sampling steps, which makes it efficient to run on the many images of a single scene. For efficiency, the diffusion process operates in the latent space of a pre-trained variational autoencoder rather than in image space. The forward noising process interpolates the target latent with Gaussian noise, $\Znoisy = (1-t)\mathbf{z}_0 + t \epsilon$, where $\mathbf{z}_0$ is the latent of the harmonized (target) image, $\epsilon\sim\mathcal{N}(0,\mathbf{I})$ is Gaussian noise, and $t$ is the timestep.

The backward process is represented by a neural network $\mathbf{v}_{\theta}$ whose weights $\theta$ are learned during fine-tuning by minimizing a rectified flow-matching loss:
\begin{equation}
    \mathcal{L}_\theta = \mathbb{E} 
    \left[ \left\| \mathbf{v}_{\theta}(\Znoisy, \Zcomp, \Zmask, c, t) -(\epsilon - \mathbf{z}_0) \right\|_2^2 \right]
\end{equation}
The network regresses the velocity $\epsilon - \mathbf{z}_0$ and is conditioned on two extra inputs: the latent of the composite image $\Zcomp$ and the latent of the object's mask $\Zmask$. These are concatenated channel-wise with the noisy latent $\Znoisy$ at the current timestep and, unlike $\Znoisy$, are not interpolated with noise. Following DEGS~\cite{DEGS}, we accommodate the extra inputs by repeating the first linear layer of the original network and copying its weights and biases to initialize the new layer. We fix the text conditioning $c$ to the empty string. The same strategy has been used for UNet-based models~\cite{zeng2024RGB-X, magar2025LightLab}.

\subsection{Data Preparation}\label{sec:relighitng_data_preparation}

To train our harmonization model, we need to form image pairs $(\Isrc, \Itgt)$, where the composite input image $\Isrc$ contains an inserted object with lighting that is inconsistent with the surrounding scene, and the ground truth target $\Itgt$ shows the same object under correct scene illumination, including consistent lighting effects such as shadows and reflections.

The source object is identified in the source image with a binary mask $\Imask$. The inconsistent input image $\Isrc$ is created by rendering an auxiliary image $\Ifg$ where the object is rendered alone with different lighting.
Then the object in image $\Ifg$ is pasted into the background image $\Ibg$ that has the original source lighting of the scene without the object (and thus without its shadows and reflections):
\begin{equation}\label{eq:data_gen}
    \Isrc = \Imask \odot \Ifg + (1 - \Imask) \odot \Ibg   
\end{equation}

By training on this image-to-image task, our model learns to map an inconsistent input image $\Isrc$, where the object appears under incorrect lighting, into a harmonized target image $\Itgt$, where the object matches the illumination of the surrounding scene. Conditioning on the object mask $\Imask$ provides the object's extent to the model.

\begin{figure}
    \centering
    \includegraphics[width=\linewidth]{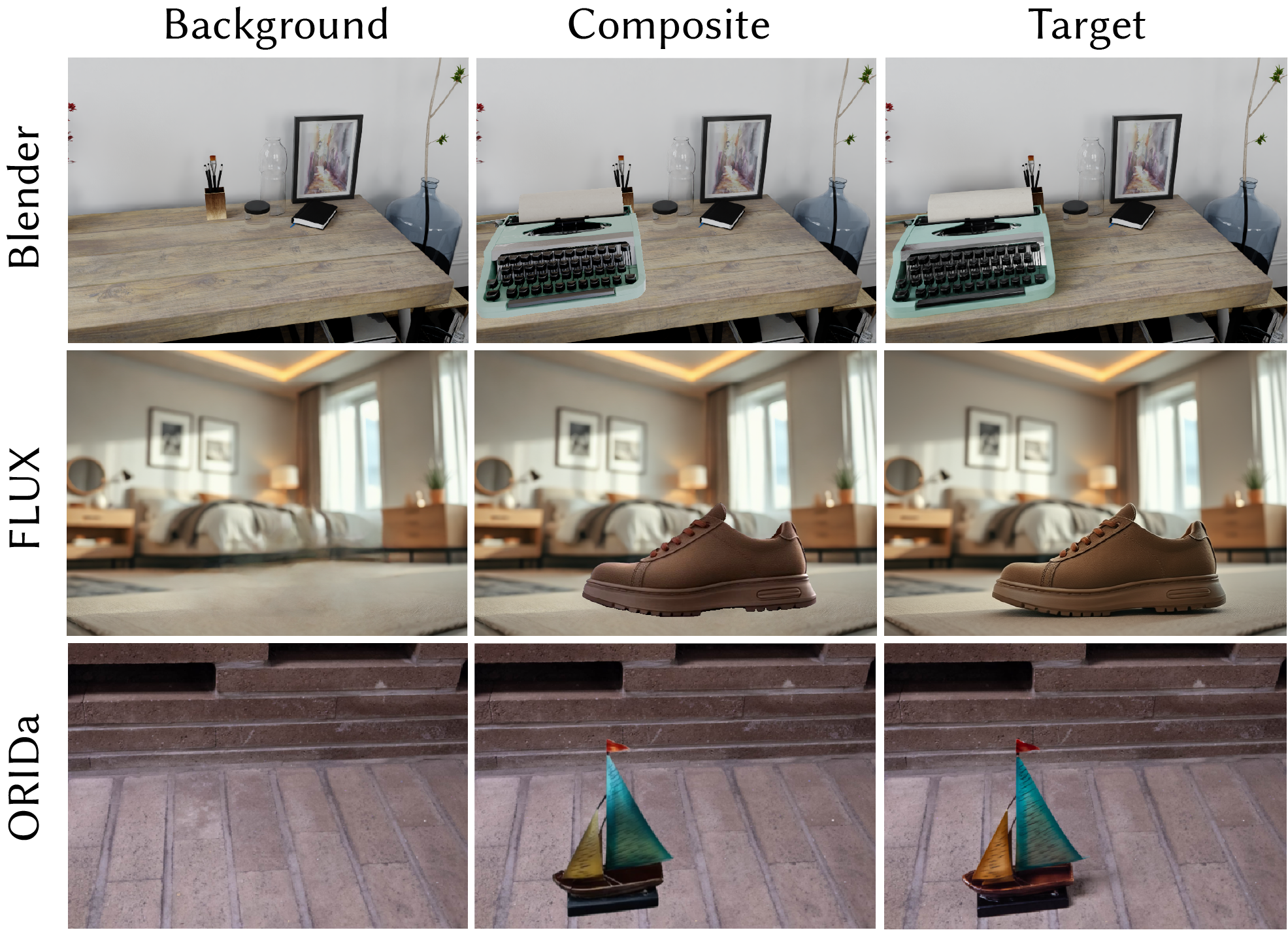}
    \caption{\textbf{Data for Harmonization Model.} Following Eq.~\ref{eq:data_gen}, we form an input image by compositing a background image of a scene with an object lit under a different illumination. The target image contains the harmonized object within the scene. During training, pairs of input and target images are sampled on-the-fly. We have three data sources: our synthetic dataset of Blender scenes, a dataset of images generated with FLUX~\cite{labs2025flux1kontextflowmatching}, and the ORIDa dataset~\cite{kim2025ORIDa}.}
    \label{fig:relight_data_gen}
\end{figure}

\subsection{Datasets}\label{sec:datasets}

\textbf{Synthetic Data.} We use 30 high-quality scenes of indoor environments including kitchens, living rooms, and offices constructed by 3D artists in Blender~\cite{blender}, and we keep two of them as a separate test set for our ablations. We manually annotate the scenes by placing cameras in suitable positions and selecting which objects will be used to produce composite images. For each camera, we define a small arc path from which 5 images are rendered. For a given camera and an object, our data generation pipeline first renders the image of the scene \textit{with} the object, which we use as a consistent target. Next, we render the scene \textit{without} the object to obtain a background.
Finally, we sample an environment map and render the object in isolation under this different illumination. To form a composite image, we simply paste the isolated object into the background image using the object mask (Eq.~\ref{eq:data_gen}). However, naively pasting the object could produce objects with large differences in exposure due to the different illumination conditions between the scene and the environment map. To address this, during training we normalize the image of the object under different illumination by the mean and standard deviation of the image of the object in the scene. We also apply this normalization to the composite images of the other two datasets.

All images are rendered in sRGB color space, using physically-based Monte Carlo path tracing (Blender Cycles~\cite{blender}) with 128 spp at resolution $704\times 496$. We automatically obtain object masks using the object ID from Blender. This dataset provides 1.9K consistent target images and over 51K inconsistent composite images from combinations of different objects and different environment maps. See Fig.~\ref{fig:relight_data_gen} (top row) for an example.

\textbf{Generated Data.} We extend our dataset of synthetic images with generated images from the text-to-image model FLUX~\cite{labs2025flux1kontextflowmatching}. We use a set of text prompts to generate high-quality images of typical objects in different indoor environments. The text prompts combine different pre-defined categories of objects, environments, lighting conditions, and points of view to produce a variety of images. Unlike the synthetic dataset, in this case we do not have direct access to an image without the object to use as background, nor to object masks. To obtain the mask of the foreground object, we use BiRefNet~\cite{zheng2024birefnet}. To address the lack of background images, we train a custom object removal network on our synthetic data. This network inputs an image of the scene with the object and a mask, and outputs a background image without the object. Most importantly, our object removal network also removes shadows and reflections caused by the object, which the harmonization model needs to learn to produce by itself. To produce the inconsistent versions of the object, we first relight the generated images using DiffusionRenderer~\cite{liang2025DiffusionRenderer} with different environment maps. Then we form the composite by segmenting the object from the relit image using the object mask, and pasting it on the background image.
With this strategy, we generate 1.6K consistent target images, and 6.5K inconsistent composite images from the different relighting results. See Fig.~\ref{fig:relight_data_gen} (middle row) for an example.

\textbf{Real Data.} 
To increase diversity in real data, we include captures from the object-centric ORIDa dataset~\cite{kim2025ORIDa}, which contains images of simple scenes with an object, its corresponding masks, and background images. However, the dataset is redundant: the exact same objects are used multiple times on the same backgrounds. For this reason, we subsample 2K images from the available 27K images to balance their contribution with respect to the other datasets. To produce the composites, we use DiffusionRenderer~\cite{liang2025DiffusionRenderer} to relight the images of the scene with the object. Then we produce the composite by segmenting the object from the relit image using the object mask, and pasting the object on the background image.
This dataset provides 2K consistent target images, and 8K inconsistent composite images from the different relighting results. See Fig.~\ref{fig:relight_data_gen} (bottom row) for an example.

\subsection{Implementation Details}

\textbf{Harmonization Model Training.} We fine-tune our harmonization model for $50K$ iterations on 4 NVIDIA H100 GPUs at resolution $704\times 496$ using Adam~\cite{kingma2014adam} with a learning rate of $3\times10^{-5}$ and a total batch size of $4$, which takes about two days. We build our training and inference code with Diffusers~\cite{von-platen-etal-2022-diffusers}. At inference, we use 4 denoising steps, which takes approximately 4 seconds on a single H100 for this resolution. Inference on new images is not restricted to the fine-tuning resolution ($704\times 496$). To handle arbitrary image resolutions, we pad them to make their resolution a multiple of $16$ (required by the underlying FLUX architecture) and crop the result.

\textbf{FLUX Dataset Generation.} To generate the FLUX dataset (Sec.~\ref{sec:datasets}), we use the original FLUX.1-schnell~\cite{labs2025flux1kontextflowmatching} with 4 inference steps and a guidance scale factor of 1. To automate the process, we use the following text prompt:

``A realistic photo of a \texttt{object} in \texttt{environment} with \texttt{lighting}, \texttt{angle}, full view of the \texttt{object}, foreground object clearly shown, resting naturally on a surface or floor, with visible contact shadow, everything in focus.''

The specific values of \texttt{object}, \texttt{environment}, \texttt{lighting}, and \texttt{angle} are taken from a set of predefined prompts. 

To obtain the background images (without shadows and reflections) needed to create the (inconsistent) composites, we train a custom object removal network since generic models~\cite{labs2025flux1kontextflowmatching,nanobanana} prompted to remove an object would occasionally change other parts of the background.
This network is fine-tuned from Stable Diffusion 2~\cite{rombachHighResolutionImageSynthesis2022} with the Diffusers codebase~\cite{von-platen-etal-2022-diffusers} using our own synthetic dataset. We train the object removal network on a single NVIDIA H100 GPU with a batch size of 8 and a learning rate of $1\times 10^{-5}$ for 260K iterations.

\section{Multi-View Consolidation}\label{sec:relight_transfer}
The input views of the target scene are independently harmonized after compositing the object. We run a post-optimization to consolidate those views into a multi-view consistent 3DGS representation (Fig.~\ref{fig:overview}). 

We initialize the representation from the composition of the object and the target scene, i.e., pasting the Gaussians of the object in the target scene. We optimize only the colors (all SH bands), keeping all other Gaussian attributes constant. 
Since the composited object is not necessarily seen from the same points of view in the source scene and in the target scene, we set the higher bands of SH to zero to reset the view-dependent effects. Otherwise, view-dependent effects from the source scene would remain from points of view that are not seen in the target scene. During post-optimization, we use a perceptual loss~\cite{zhang2018lpips} because its focus on higher-level semantic similarity, rather than pixel-wise photometric error, makes it more robust to potential inconsistencies across views~\cite{gaoCAT3DCreateAnything2024}. 

Our implementation is based on the original 3DGS codebase~\cite{kerbl_3Dgaussians_2023}, which includes the training speed acceleration from Taming-3DGS~\cite{mallick2024taming}. We run our post-optimization for 10K iterations with the default parameters of 3DGS. We also disable densification and opacity reset since the geometry is already consolidated.

In Section~\ref{sec:ablations}, we discuss and evaluate other consolidation strategies based on iterative dataset updates~\cite{haque2023instruct}, a popular approach for radiance field editing, where the diffusion model is run multiple times on renderings from previously consolidated views, gradually converging to a multi-view consistent radiance field. Unlike some editing methods, which need to harmonize the color and the underlying geometry of the radiance field, we only need to harmonize its color. For this reason, it is easier for our post-optimization to rapidly converge to a multi-view consistent solution, without multiple updates of the dataset. The benefit of our post-optimization is that the harmonization model only needs to be run \textit{once} per view.
\section{Evaluation}
We present qualitative evaluation on real scenes, quantitative evaluation using synthetic scenes where ground truth is available, and also ablations to better understand the importance of each component of our method.

\begin{figure*}
    \centering
    \includegraphics[width=\linewidth]{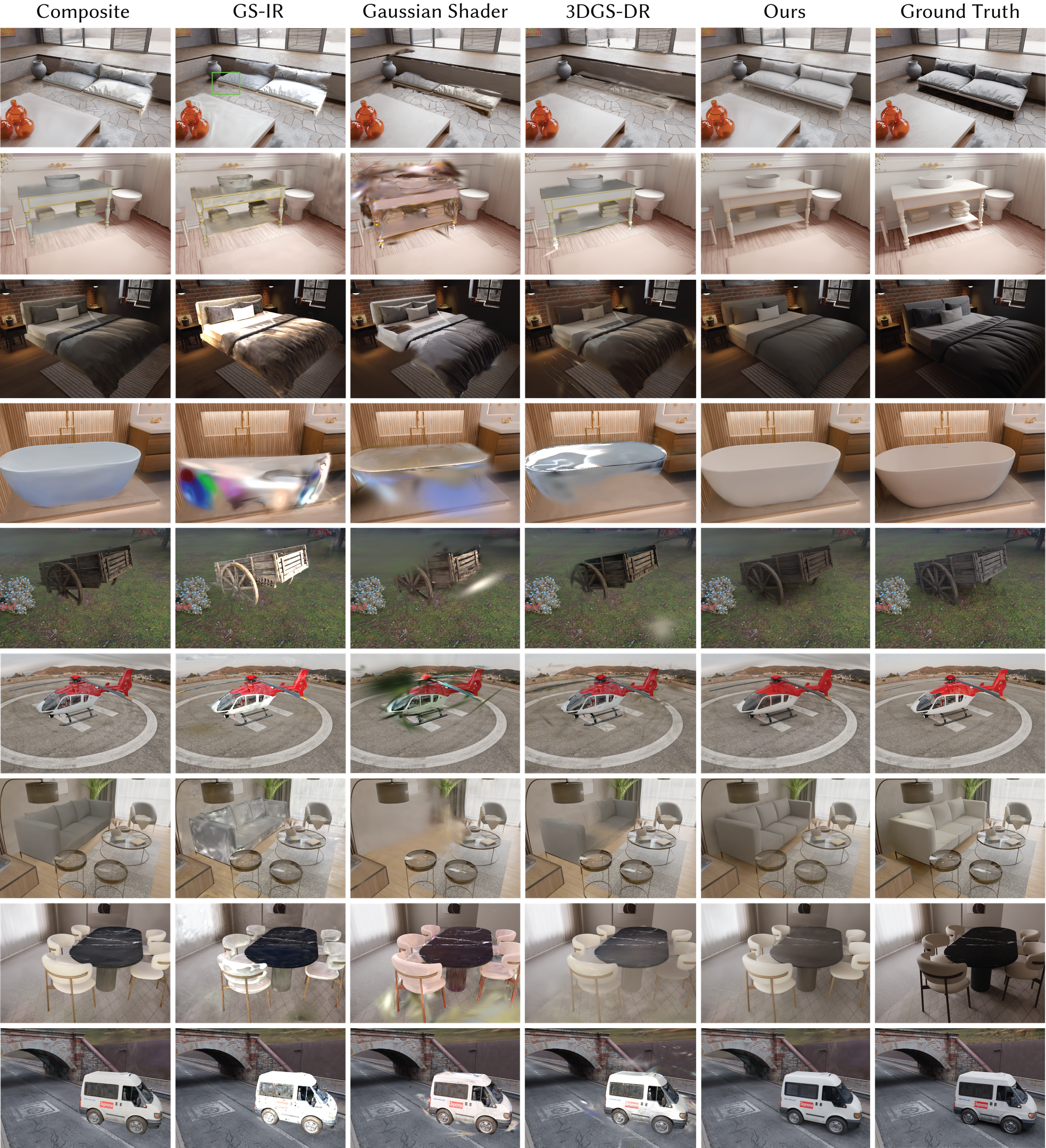}
    \caption{\textbf{Qualitative 3D Results.} We provide nine high-quality synthetic scenes with ground truth for numerical comparisons. From left to right: Composite, GS-IR~\cite{liangGSIR3DGaussian2024}, Gaussian Shader~\cite{jiangGaussianShader3DGaussian2024}, 3DGS-DR~\cite{ye3DGaussianSplatting2024}, Ours, and Ground Truth. }
    \label{fig:3d_comparison_synthetic}
\end{figure*}

\textbf{Datasets.} We demonstrate our method on a variety of scenes taken from 
Mip-NeRF-360~\cite{barron2022mip-nerf360}, Zip-NeRF~\cite{barron2023zip}, LERF~\cite{kerr_lerf_2023}, DL3DV~\cite{ling2024dl3dv}, Ref-NeRF~\cite{verbin2021ref-nerf}, and Tanks and Temples~\cite{TanksAndTemplates2017Knapitsch}. We include some synthetic objects for insertion (plant, desk, locomotive) from BlenderKit, reconstructed with 3DGS.
For quantitative results with a known ground truth, we provide nine extra high-quality synthetic scenes (see Fig.~\ref{fig:3d_comparison_synthetic}). We render images of these scenes with Blender Cycles~\cite{blender}, which are then used by the 3DGS reconstruction pipeline.

\begin{figure*}
    \centering
    \includegraphics[width=\linewidth]{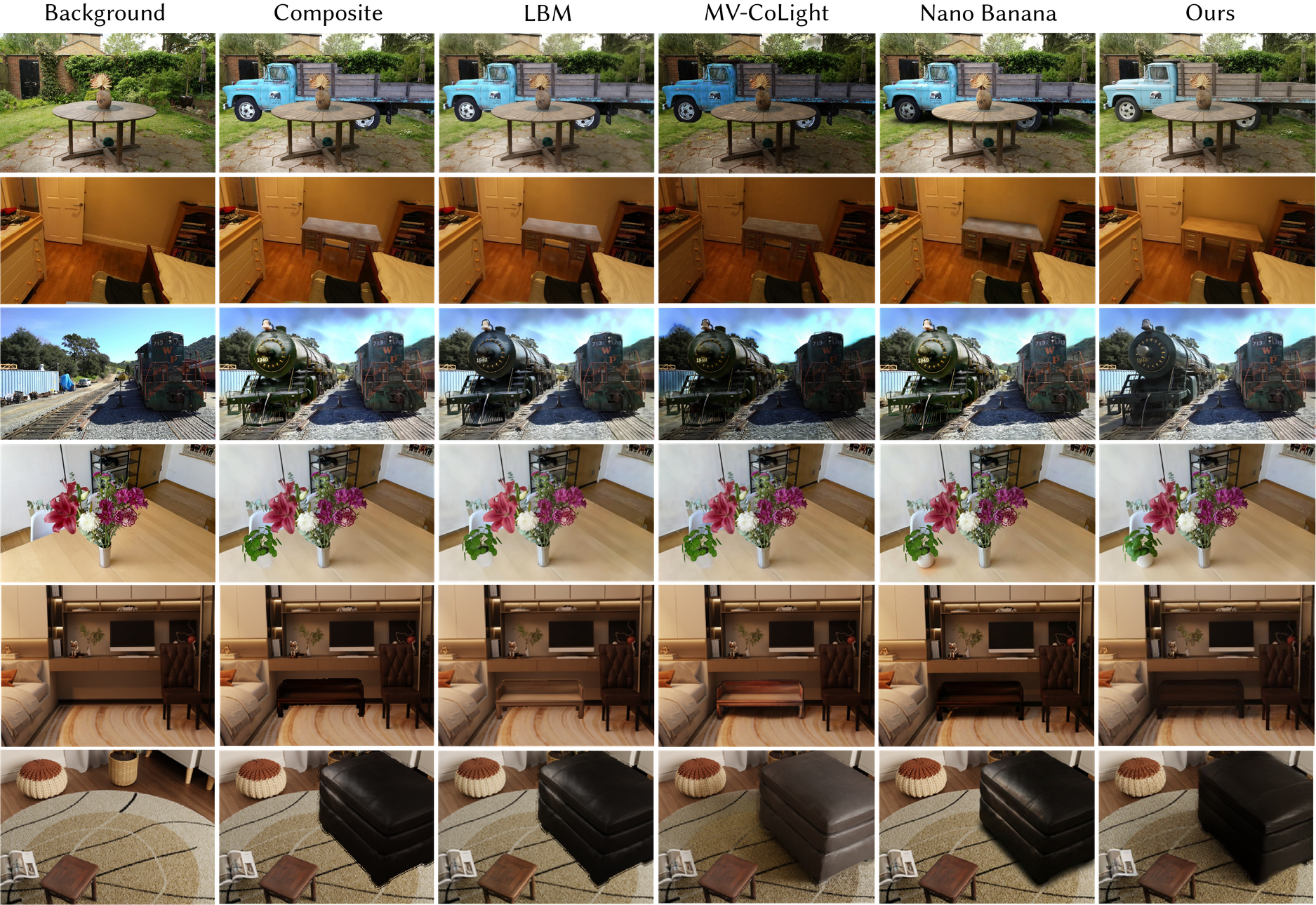}
    \caption{\textbf{Qualitative 2D Results.} From left to right: Background, Composite, LBM~\cite{chadebec2025lbm}, MV-CoLight~\cite{renMVCoLightEfficientObject2025a}, Nano Banana~\cite{nanobanana}, and Ours. From top to bottom, the composited scenes are: \textit{(i)} Truck from Tanks and Temples into Garden from Mip-NeRF-360, \textit{(ii)} Desk from BlenderKit into London Bedroom from Zip-NeRF, \textit{(iii)} Locomotive from BlenderKit into Train from Tanks and Temples, \textit{(iv)} Plant from BlenderKit into Bouquet from LERF. The last two scenes, \textit{(v)} and \textit{(vi)}, are taken from the DTC-MultiLight dataset (MV-CoLight)~\cite{renMVCoLightEfficientObject2025a}.}
    \label{fig:2d_comparison}
\end{figure*}

\textbf{Qualitative 2D Evaluation.} For image harmonization, we compare our harmonization model with LBM~\cite{chadebec2025lbm}, MV-CoLight~\cite{renMVCoLightEfficientObject2025a}, and Nano Banana~\cite{nanobanana} in Fig.~\ref{fig:2d_comparison}. This allows us to assess both the overall quality of the generated images and the model’s ability to generalize across different scenarios while maintaining consistent object integration and illumination adaptation. The limitations of these methods motivate training our own harmonization model.

Nano Banana~\cite{nanobanana} sometimes exaggerates the shadows (e.g., first row, where the shadow of the truck is too strong for the scene), has color bleeding (fourth row, where the base of the pot has a green color bleeding effect) and is overall missing relighting and directional cues for providing realistic harmonization (e.g., third row, where the locomotive is not properly harmonized). To use Nano Banana as a harmonization model, we provide a prompt consisting of the composite image, the object mask, and the following text: 

``I copy-pasted an object from one photograph to the other. I will give you the photograph and the mask of the copied object, white is the object and black is the background. Edit the photograph such that the copied object looks realistically composited in the environment. Create realistic contact shadows, ambient occlusion, global illumination, accurate lighting match, seamless blending etc. This is crucial: Do not change anything else other than the composited object.''

LBM~\cite{chadebec2025lbm} is able to modify the object's lighting only for some images (truck, locomotive), but in all cases fails to produce realistic harmonization due to lack of shadows. 

MV-CoLight~\cite{renMVCoLightEfficientObject2025a} only produces satisfactory results for its in-distribution scenes from the DTC-MultiLight dataset (last two rows of Fig.~\ref{fig:2d_comparison}), but fails to correctly harmonize out-of-distribution images of general real scenes.

\textbf{Qualitative 3D Evaluation.} To our knowledge, no prior method specifically addresses cross-scene 3DGS object transfer with lighting harmonization. Inverse rendering methods, which decompose scenes into intrinsic components to enable relighting, are therefore the most directly comparable approaches for this task, and we compare our method against Gaussian Shader~\cite{jiangGaussianShader3DGaussian2024}, GS-IR~\cite{liangGSIR3DGaussian2024}, and 3DGS-DR~\cite{ye3DGaussianSplatting2024}, which we have extended to segment and insert an object from a source to a target scene. For other related methods, code is not publicly available (\cite{renGauUpdateNewObject2025, zhuRelightingScenesObject2024}). 

\begin{figure*}
    \centering
    \includegraphics[width=\linewidth]{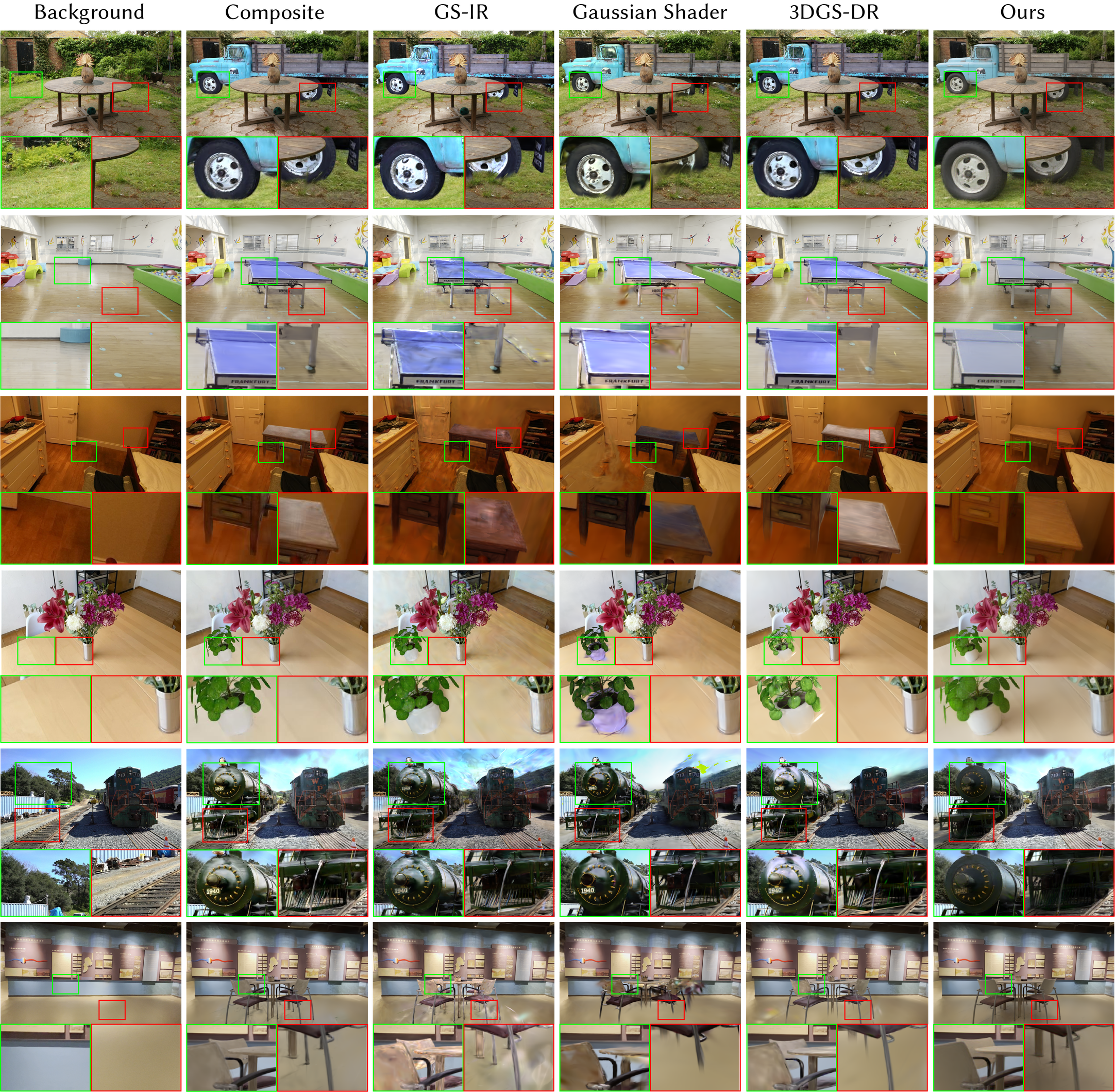}
    \caption{\textbf{Qualitative 3D Results.} From left to right: Background, Composite, GS-IR~\cite{liangGSIR3DGaussian2024}, Gaussian Shader~\cite{jiangGaussianShader3DGaussian2024}, 3DGS-DR~\cite{ye3DGaussianSplatting2024}, and Ours. From top to bottom, the composited scenes are: \textit{(i)} Truck from Tanks and Temples into Garden from Mip-NeRF-360, \textit{(ii)} 508850 into 4cea29, both from DL3DV, \textit{(iii)} Desk from BlenderKit into London Bedroom from Zip-NeRF, \textit{(iv)} Plant from BlenderKit into Bouquet from LERF, \textit{(v)} Locomotive from BlenderKit into Train from Tanks and Temples, and \textit{(vi)} d11d95 into 83d5f2, both from DL3DV. See the zoomed-in regions (green and red) for more details.}
    \label{fig:3d_comparison_real_synthetic_mix}
\end{figure*}

We show different insertion results in Fig.~\ref{fig:3d_comparison_real_synthetic_mix} on real scenes and Fig.~\ref{fig:3d_comparison_synthetic} on synthetic scenes, demonstrating that our method effectively learns to harmonize the illumination of the object with its surroundings while seamlessly integrating it into the scene with consistent shadows and reflections. We evaluate our model on a wide range of scenes (both indoors and outdoors) and objects, covering diverse geometries, materials, and lighting conditions. 
\newpage
GS-IR~\cite{liangGSIR3DGaussian2024} tends to produce overly saturated results on the object's surface, especially for white objects, as shown in Fig.~\ref{fig:3d_comparison_synthetic} (e.g., the bed in the third row). Also, it struggles to adapt the overall lighting of the object to match the target scene, and it lacks contact shadows.

Gaussian Shader~\cite{jiangGaussianShader3DGaussian2024} produces more natural results without saturation on white objects but, similar to GS-IR, it does not adapt the lighting of the object and lacks shadows, resulting in unrealistic scenes.

3DGS-DR~\cite{ye3DGaussianSplatting2024} struggles with the same issue: lack of proper harmonization and shadows, making the final result look unrealistic.

Overall, previous methods fail to correctly capture shadows and illumination conditions. In contrast, our method produces realistic object harmonization and creates proper shadows, making the insertion look natural. Fig.~\ref{fig:3d_comparison_synthetic} contains cases with pronounced lighting mismatches: the composite sofa has a strong uneven illumination which our model correctly harmonizes; the bed has a strong highlight from the source scene which is also correctly handled. In Fig.~\ref{fig:3d_comparison_real_synthetic_mix}, our method corrects the lighting in the front of the locomotive (fifth row), which is not achieved by other methods. 

\textbf{Quantitative 3D Evaluation.} We evaluate our method on nine high-quality synthetic scenes: six indoors and three outdoors (Fig.~\ref{fig:3d_comparison_synthetic}). In Table~\ref{tab:quantitatve_comparison_synthetic}, we report PSNR, SSIM, and LPIPS~\cite{zhang2018lpips} to measure reconstruction quality with respect to the ground truth. Since our harmonization model is at its core a generative model, it produces a \textit{plausible} harmonization, which does not necessarily match the ground truth at a pixel level. To measure how similar the harmonized images are with respect to the ground truth at a distribution level, we also compute FID~\cite{heusel2017fid} and KID~\cite{binkowski2018kid}, which are typically used to evaluate the realism of generated images. Our method outperforms previous alternatives by a wide margin across all metrics. 

\begin{figure}
    \centering
    \includegraphics[width=\linewidth]{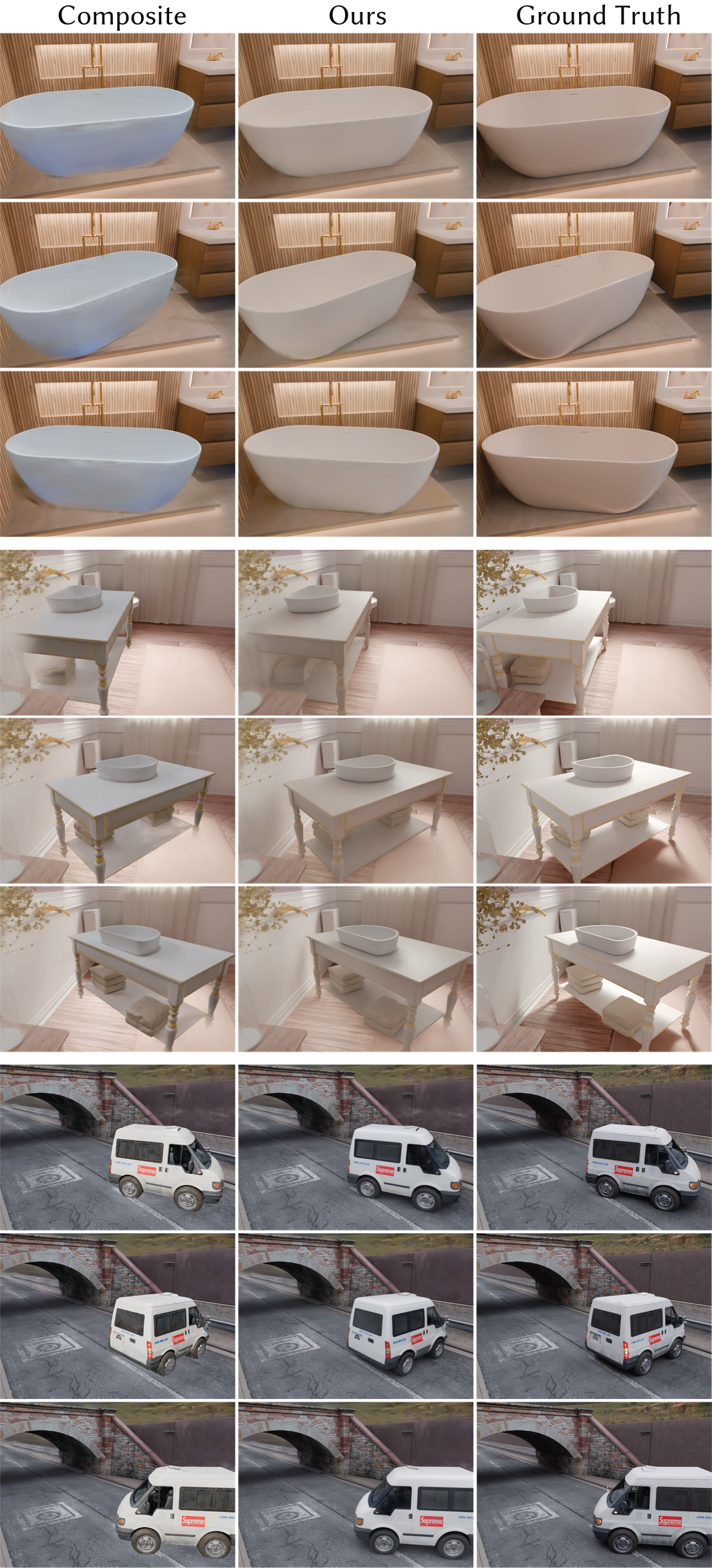}
    \caption{\textbf{Illumination Consistency.} Results of our harmonization method on three synthetic scenes with an inserted object at three different rotations. Our method produces consistent multi-view illumination across all configurations.}
    \label{fig:rotation_exp}
\end{figure}

\newpage
\textbf{Illumination Consistency.} To evaluate the consistency of our illumination harmonization across different object configurations, we conduct an experiment on three synthetic scenes where we insert an object into each scene and rotate it across three poses $(P_0,P_1,P_2)$. We run our harmonization model independently on each composite scene and then we run the post-optimization and compare with the GT for each position (we hold out every 8th image as a test image, following Mip-NeRF-360~\cite{barron2022mip-nerf360}), obtaining similar metrics across scenes (Table~\ref{tab:rotation_experiment}).
By comparing the harmonization and post-optimization with the ground truth (which is by definition multi-view consistent) in novel views, we are measuring the multi-view consistency of our approach. 
We show qualitative results in Fig.~\ref{fig:rotation_exp}, where we observe a consistent illumination for the different positions of the objects in the scene.
While a strong cast shadow is not produced for the bathroom, our model generates a plausible harmonization across views in all cases, including soft shadows on the walls.

\begin{table}
  \centering
    \caption{\textbf{Quantitative 3D Evaluation.} We report metrics on nine synthetic scenes (Fig.~\ref{fig:3d_comparison_synthetic}), comparing our method against three inverse rendering techniques extended for object transfer across scenes: Gaussian Shader~\cite{jiangGaussianShader3DGaussian2024}, GS-IR~\cite{liangGSIR3DGaussian2024}, and 3DGS-DR~\cite{ye3DGaussianSplatting2024}.}
  \begin{tabular}{l|ccccc}
   & PSNR$\uparrow$ & SSIM$\uparrow$ & LPIPS$\downarrow$ & FID$\downarrow$ & KID$\downarrow$ \\ \hline
  Ours & \cellcolor{tabfirst}22.53 & \cellcolor{tabfirst}0.840 & \cellcolor{tabfirst}0.214 & \cellcolor{tabfirst}55.86 & \cellcolor{tabfirst}0.0413 \\
  GShader & \cellcolor{tabthird}18.66 & 0.799 & 0.272 & 152.62 & 0.1742 \\
  GS-IR & 17.83 & \cellcolor{tabthird}0.803 & \cellcolor{tabthird}0.247 & \cellcolor{tabsecond}110.96 & \cellcolor{tabsecond}0.1143 \\
  3DGS-DR & \cellcolor{tabsecond}19.65 & \cellcolor{tabsecond}0.814 & \cellcolor{tabsecond}0.245 & \cellcolor{tabthird}124.07 & \cellcolor{tabthird}0.1448 \\
  \end{tabular}
  \label{tab:quantitatve_comparison_synthetic}
\end{table}

\begin{table}
    \centering
    \caption{\textbf{Illumination Consistency.} We evaluate the illumination consistency of three scenes where an object has been inserted and rotated in-place in three different positions: $P_0, P_1, P_2$ (see Fig.~\ref{fig:rotation_exp}). We report average results across the scenes for each position.}
    \begin{tabular}{l|ccccc}
     & PSNR$\uparrow$ & SSIM$\uparrow$ & LPIPS$\downarrow$ & FID$\downarrow$ & KID$\downarrow$ \\
    \hline
    $P_0$ & 25.14 & 0.9135 & 0.1553 & 46.41 & 0.0161 \\
    $P_1$ & 24.85 & 0.9117 & 0.1589 & 38.34 & 0.0118 \\
    $P_2$ & 24.16 & 0.9097 & 0.1639 & 45.78 & 0.0160 \\
    \end{tabular}
    \label{tab:rotation_experiment}
\end{table}

\subsection{Ablations}\label{sec:ablations}

\textbf{Data for Harmonization Model.} We assess the impact of each one of our three data sources (Blender, FLUX, ORIDa) by training a harmonization model without each one of them, and evaluating the resulting model on a separate test split of each dataset. Our results, summarized in Table~\ref{tab:data_ablations}, show that our carefully curated Blender dataset has the most impact, significantly improving metrics. The FLUX dataset also has an important but lesser impact. The ORIDa dataset has the least impact overall: including it in the training process improves PSNR and FID slightly but worsens SSIM and LPIPS. We believe this is due to the limited diversity of the dataset, which consists of a short list of simple objects and backgrounds.

\begin{table*}
    \centering
    \caption{\textbf{Dataset Ablation.} We analyze the impact on our diffusion model of each one of the three data sources: synthetic data from Blender, generated images with FLUX~\cite{labs2025flux1kontextflowmatching}, and real images from ORIDa~\cite{kim2025ORIDa}. For each dataset, metrics are computed in a separate test set not seen during training.}
    \resizebox{\textwidth}{!}{
    \begin{tabular}{l|cccc|cccc|cccc|cccc}
     & \multicolumn{4}{c|}{Blender} & \multicolumn{4}{c|}{FLUX} & \multicolumn{4}{c|}{ORIDa} & \multicolumn{4}{c}{Average} \\
     & PSNR$\uparrow$ & SSIM$\uparrow$ & LPIPS$\downarrow$ & FID$\downarrow$ & PSNR$\uparrow$ & SSIM$\uparrow$ & LPIPS$\downarrow$ & FID$\downarrow$ & PSNR$\uparrow$ & SSIM$\uparrow$ & LPIPS$\downarrow$ & FID$\downarrow$ & PSNR$\uparrow$ & SSIM$\uparrow$ & LPIPS$\downarrow$ & FID$\downarrow$ \\ \hline

Ours    &  \cellcolor{tabfirst}27.08 & \cellcolor{tabsecond}0.893 & \cellcolor{tabsecond}0.124 &  \cellcolor{tabfirst}33.86 &  \cellcolor{tabfirst}28.34 &  \cellcolor{tabfirst}0.892 &  \cellcolor{tabthird}0.108 & \cellcolor{tabsecond}25.13 & \cellcolor{tabsecond}30.08 &  \cellcolor{tabthird}0.865 &                      0.170 &  \cellcolor{tabthird}74.64 &  \cellcolor{tabfirst}28.50 & \cellcolor{tabsecond}0.883 & \cellcolor{tabsecond}0.134 &  \cellcolor{tabfirst}44.54 \\
w/o Blender &                      25.69 &                      0.870 &                      0.164 &                      47.78 &  \cellcolor{tabthird}28.28 & \cellcolor{tabsecond}0.890 & \cellcolor{tabsecond}0.106 &  \cellcolor{tabthird}26.90 &                      29.71 &                      0.863 & \cellcolor{tabsecond}0.160 &                      86.49 &  \cellcolor{tabthird}27.89 &  \cellcolor{tabthird}0.874 &  \cellcolor{tabthird}0.143 &                      53.72 \\
w/o Flux    & \cellcolor{tabsecond}26.98 &  \cellcolor{tabfirst}0.895 &  \cellcolor{tabfirst}0.122 & \cellcolor{tabsecond}34.64 &                      24.70 &  \cellcolor{tabthird}0.852 &                      0.147 &                      46.23 &  \cellcolor{tabfirst}30.41 & \cellcolor{tabsecond}0.868 &  \cellcolor{tabthird}0.163 &  \cellcolor{tabfirst}70.19 &                      27.36 &                      0.872 &                      0.144 &  \cellcolor{tabthird}50.36 \\
w/o Orida   &  \cellcolor{tabthird}26.93 &  \cellcolor{tabthird}0.890 &  \cellcolor{tabthird}0.130 &  \cellcolor{tabthird}35.93 & \cellcolor{tabsecond}28.29 &  \cellcolor{tabfirst}0.892 &  \cellcolor{tabfirst}0.104 &  \cellcolor{tabfirst}24.66 &  \cellcolor{tabthird}29.92 &  \cellcolor{tabfirst}0.887 &  \cellcolor{tabfirst}0.126 & \cellcolor{tabsecond}73.46 & \cellcolor{tabsecond}28.38 &  \cellcolor{tabfirst}0.890 &  \cellcolor{tabfirst}0.120 & \cellcolor{tabsecond}44.68

    \end{tabular}}
    \label{tab:data_ablations}
\end{table*}

\newpage
\textbf{Iterative Dataset Update.}
We show that iterative dataset updates~\cite{haque2023instruct} are not necessary for object transfer across radiance fields: a simpler approach with a single update suffices to obtain realistic transfer results. 
An iterative approach alternates two steps: 1) applying the harmonization model to renders of already consolidated views (with decreasing values of starting noise), and 2) consolidating the current dataset of individually harmonized views. The dataset of views of the target scene is thus gradually updated, and the 3DGS representation converges to a multi-view consistent representation.

For all noise levels we use the composite image as conditioning for the harmonization model; early experiments using the subsequent renders as conditioning produced blurry results.
We also try two different dataset update schedules: a linear and a geometric progression. In both cases, the first dataset update is done running our harmonization model starting from complete noise. But subsequent updates start from \textit{lower} levels of noise by encoding the rendering from a given point of view in latent space and adding noise to it (e.g., adding noise at 80\%). The diffusion process starts from this level of noise and proceeds until a clean image is obtained. Later updates use lower levels of noise, thus gradually converging to a multi-view consistent result.

\begin{figure}
    \centering
    \includegraphics[width=\linewidth]{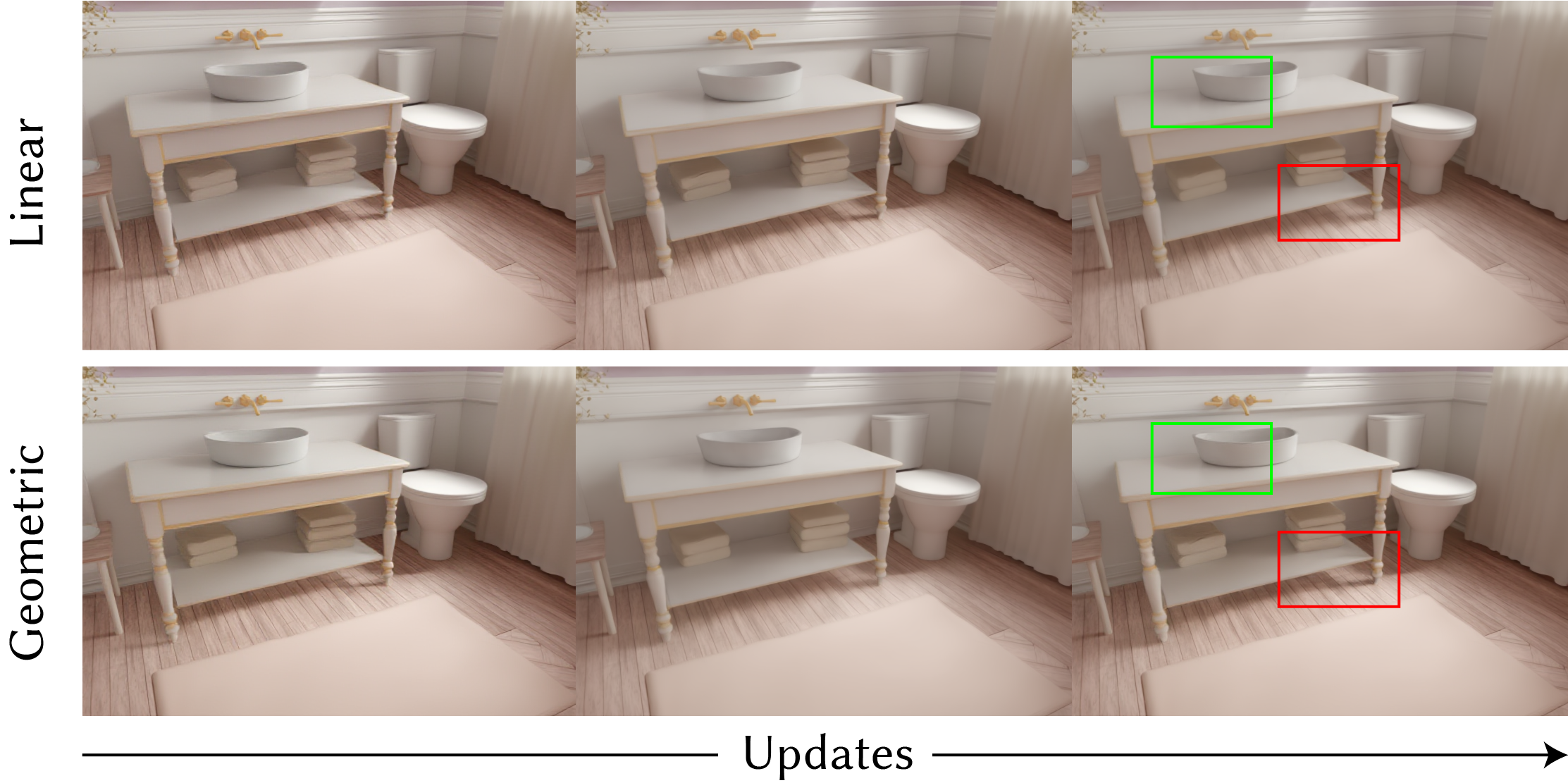}
    \caption{\textbf{Iterative Dataset Update Ablation.} We experimented with two iterative dataset update schedules for the consolidation: linear from $1.0$ to $0.1$ and geometric from $1.0$ to $0.1$. In this example we perform five dataset updates (only updates 1, 3, and 5 are shown here). But we still found that a single consolidation produced satisfactory results without having to update the dataset multiple times (see Fig.~\ref{fig:3d_comparison_synthetic}).}
    \label{fig:iterative_updates}
\end{figure}

We show a qualitative example (Fig.~\ref{fig:iterative_updates}) comparing the geometric and the linear progression. But we opt for the more efficient solution of doing only one consolidation step, since it produces better results without having to update the dataset multiple times, thus being faster, as shown in Table~\ref{tab:iterative_updates}.

\begin{table}
  \centering
    \caption{\textbf{Iterative Dataset Update Ablation.} We analyze the impact of using iterative dataset updates~\cite{haque2023instruct} with a linear and a geometric noise schedule with five dataset updates.}
  \resizebox{\columnwidth}{!}{%
  \begin{tabular}{l|cccccc}
   & Post-opt $\downarrow$ & PSNR$\uparrow$ & SSIM$\uparrow$ & LPIPS$\downarrow$ & FID$\downarrow$ & KID$\downarrow$ \\ \hline
  Ours & \cellcolor{tabfirst}15m30s & \cellcolor{tabfirst}22.53 & \cellcolor{tabfirst}0.840 & \cellcolor{tabfirst}0.214 & \cellcolor{tabfirst}55.86 & \cellcolor{tabfirst}0.0413 \\
  w/Linear & \cellcolor{tabsecond}1h8m & \cellcolor{tabsecond}21.79 & \cellcolor{tabsecond}0.826 & \cellcolor{tabsecond}0.225 & \cellcolor{tabsecond}59.46 & \cellcolor{tabsecond}0.0462 \\
  w/Geometric & \cellcolor{tabsecond}1h8m & \cellcolor{tabthird}21.55 & \cellcolor{tabthird}0.820 & \cellcolor{tabthird}0.230 & \cellcolor{tabthird}60.47 & \cellcolor{tabthird}0.0477 \\
  \end{tabular}}
  \label{tab:iterative_updates}
\end{table}

\textbf{Computational Cost.} We run the single-view harmonization and the multi-view consolidation on a single NVIDIA H100 GPU. On our synthetic scenes of resolution $704\times 496$ consisting of $200$ images, the reconstruction of the source and target scenes takes 2m30s for each one. Training the binary features for segmentation for 10K iterations takes 40s. For the multi-view consolidation, running our harmonization model on the $200$ composite images takes 12m30s, and the 3DGS post-optimization takes 3m.

\begin{figure}
    \centering
    \includegraphics[width=\linewidth]{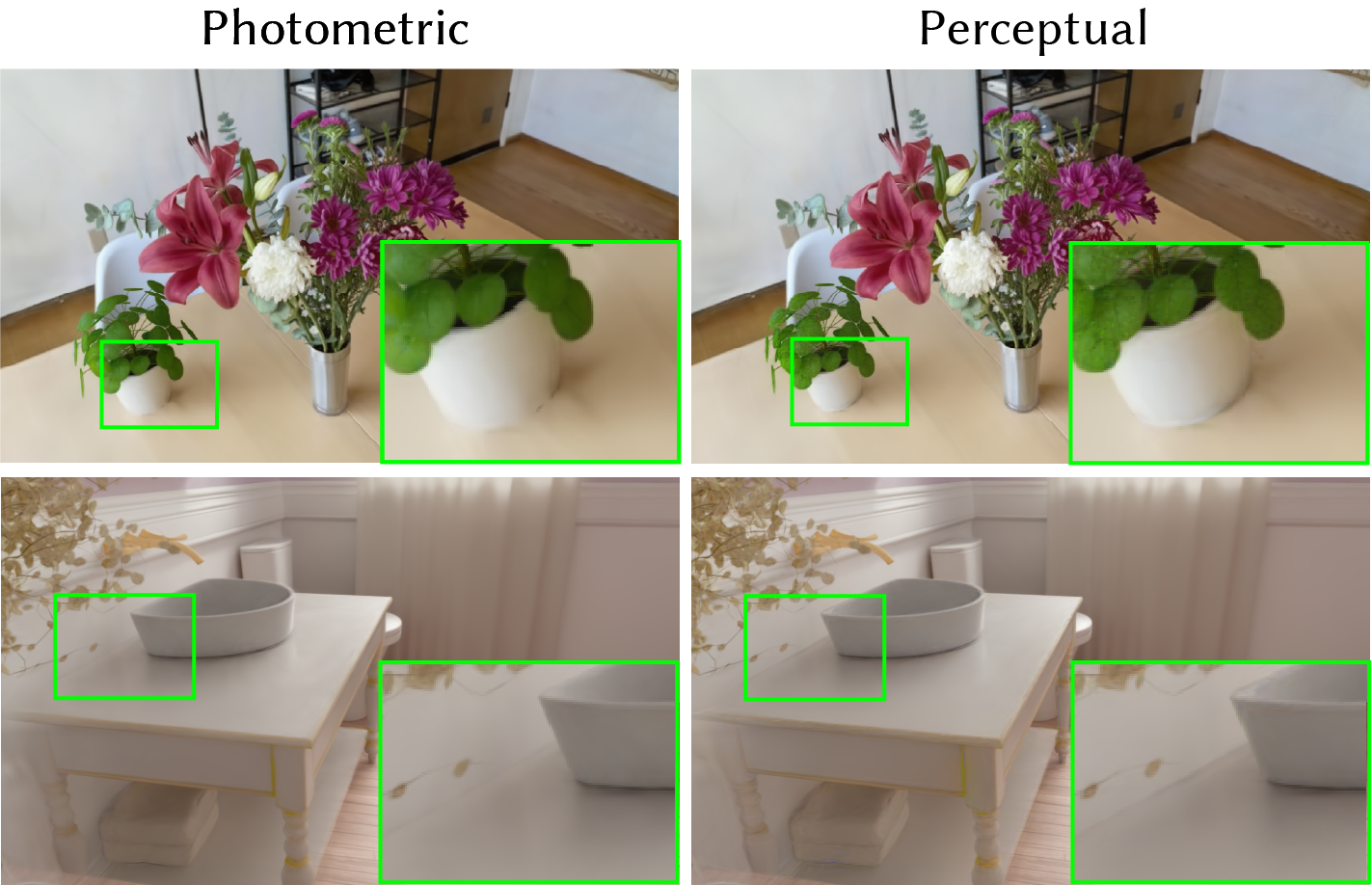}
    \caption{\textbf{Photometric vs Perceptual Loss Ablation.} For 3DGS post-optimization, a perceptual loss better preserves details and sharp edges. See zoomed-in regions (green) for details.}
    \label{fig:photometric_vs_perceptual}
\end{figure}

\break
\textbf{Perceptual vs Photometric Loss.} We use a perceptual loss~\cite{zhang2018lpips} for the 3DGS post-optimization. We observed improved results when compared to photometric error (L1/SSIM), which creates some small artifacts and more blurry results (qualitative examples in Fig.~\ref{fig:photometric_vs_perceptual}). On the other hand, post-optimization using a perceptual loss is slower than using a photometric loss. On our synthetic scenes of resolution $704\times 496$ consisting of 200 images, using a single NVIDIA H100 GPU, the post-optimization takes 1m with L1/SSIM and 3m with the perceptual loss.

\subsection{Limitations \& Future Work}
\begin{figure}
    \centering
    \includegraphics[width=\linewidth]{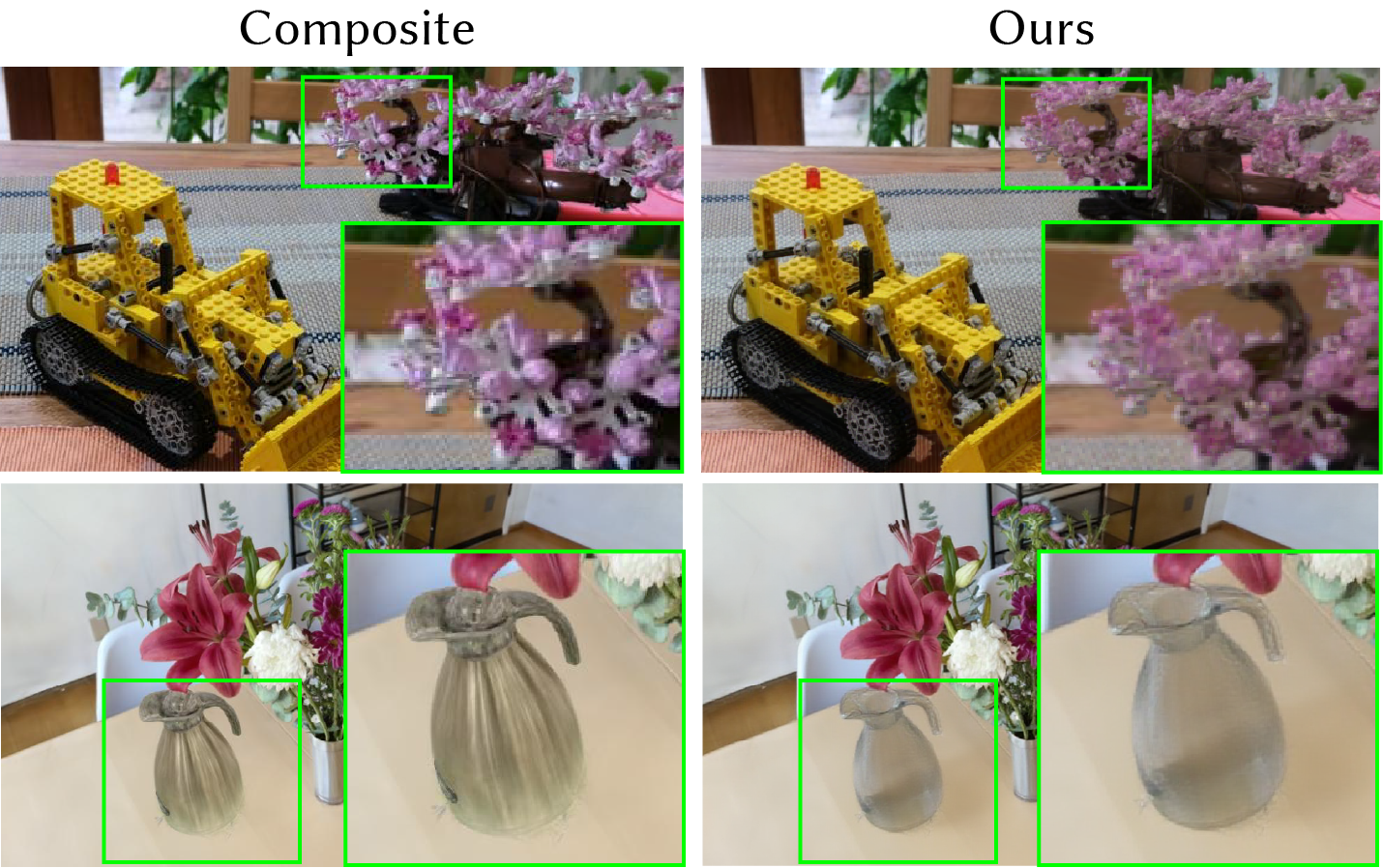}
    \caption{\textbf{Limitations.} \textit{(top)} Bonsai into Kitchen, both from Mip-NeRF-360~\cite{barron2022mip-nerf360}. Contains block-like artifacts and blur due to very small details with high-frequency content. \textit{(bottom)} Kettle from NERO~\cite{liuNeRONeuralGeometry2023} into Bouquet from LERF~\cite{kerr_lerf_2023}. Despite the harmonization looking realistic, the appearance of the object changes from a metallic surface to glass.}
    \label{fig:failure}
\end{figure}

Our method is not without limitations. The variational autoencoder inherited from FLUX introduces some artifacts for objects with very high-frequency details, where the harmonized image has blur or block-like artifacts. Also, for some objects our harmonization network changes the appearance too much, altering the material properties. We show examples of these cases in Fig.~\ref{fig:failure}.

Our method has no guarantee of absolute accuracy in the harmonization, only plausible results. Further scaling of the synthetic dataset with more scenes as well as capturing high-quality real data on complex scenes are promising directions to improve generalization.
Finally, addressing the multi-view consistency with recent video models~\cite{kong2024hunyuanvideo, wan2025, nvidiaCosmosWorldFoundation2025} constitutes another interesting avenue for future research.

\section{Conclusions}
We presented DOT3D, a method that transfers 3D objects from one scene into another, maintaining consistent lighting including shadows and reflections. Our key contribution is a 3D lighting-consistent object transfer solution based on 3DGS, which first uses a fine-tuned diffusion model to harmonize the lighting of individual images after composition. These corrected views are consolidated in 3D via a post-optimization step, creating a full 3DGS scene with multi-view consistent lighting. We demonstrated our method on a variety of scenes from widely used datasets, and also provided quantitative results on high-quality synthetic scenes with ground truth. Our solution enables asset reuse across different scenes, ensuring easy and realistic integration in new environments, with applications spanning many fields.

\noindent\textbf{Acknowledgments} 

\noindent This work was funded by the European Union, European Research Council (ERC) Advanced Grants NERPHYS, 101141721 \url{https://project.inria.fr/nerphys} and EXPLORER, 101097259 \url{https://cordis.europa.eu/project/id/101097259}. Views and opinions expressed are however those of the author(s) only and do not necessarily reflect those of the European Union or the European Research Council. Neither the European Union nor the granting authority can be held responsible for them. Experiments presented in this paper were carried out using the Grid’5000 testbed, supported by a scientific interest group hosted by Inria and including CNRS, RENATER and several Universities as well as other organizations (see \url{https://www.grid5000.fr}). The authors would also like to thank Adobe and NVIDIA for software and hardware donations.


\printbibliography             

\end{document}